\documentclass[9pt]{extarticle}
\usepackage{spconf}
\usepackage{amsthm}
\usepackage{amsfonts}
\usepackage{amssymb}
\usepackage{mathrsfs}
\usepackage{mathtools}
\usepackage[english]{babel}
\usepackage{float}
\usepackage[toc,page]{appendix}
\usepackage[pdftex]{graphicx}
\usepackage{epstopdf}
\usepackage{amsmath}
\usepackage{color}
\usepackage{multirow}
\usepackage{graphicx}
\usepackage{bm}
\usepackage{balance}
\usepackage{microtype}
\usepackage[caption=false,font=footnotesize]{subfig}

\usepackage{url}

\usepackage{textcomp}

\usepackage{cite}
\usepackage[]{algorithmicx}
\usepackage{algpseudocode,algorithm,lipsum} 

\usepackage{color}
\usepackage[table]{xcolor}% http://ctan.org/pkg/xcolor

\DeclareMathOperator*{\minimize}{minimize} % thin space, limits underneath in displays
\DeclareMathOperator*{\subjectto}{subject \hspace{3pt} to } % thin space, limits underneath in 

\title{FEDERATED DROPOUT LEARNING FOR HYBRID BEAMFORMING WITH SPATIAL PATH INDEX MODULATION IN MULTI-USER MMWAVE-MIMO SYSTEMS   }
\name{Ahmet~M.~Elbir$^{\dagger}$, Sinem Coleri$^\dagger$ and Kumar Vijay Mishra$^+$ %\thanks{Thanks to XYZ agency for funding.}
}
\address{	$^{\dagger}$Department of Electrical and Electronics Engineering, Ko\c{c} University, Istanbul, Turkey  \\
	$^+$United States CCDC Army Research Laboratory, Adelphi, MD 20783 USA \\
	\thanks{		S. Coleri acknowledges the support of the Scientific and Technological Research Council of Turkey (TUBITAK) EU CHIST-ERA grant 119E350. A. M. Elbir acknowledges the support of TUBITAK.}\\
}

\begin{document}
	% Reduce spacing above and below equations
	\setlength{\abovedisplayskip}{3pt}
	\setlength{\belowdisplayskip}{3pt}
	
	\maketitle
	
	\begin{abstract}
		Millimeter wave multiple-input multiple-output (mmWave-MIMO) systems with small number of radio-frequency (RF) chains have limited multiplexing gain. Spatial path index modulation (SPIM)  is helpful in improving this gain by utilizing additional signal bits modulated by the indices of spatial paths.	In this paper, we introduce model-based and model-free frameworks for beamformer design in multi-user SPIM-MIMO systems. We first design the beamformers via model-based manifold optimization algorithm. Then, we leverage federated learning (FL) with dropout learning (DL) to train a learning model on the local dataset of users, who estimate the beamformers by feeding the model with their channel data. The DL randomly selects different set of model parameters during training, thereby further reducing the transmission overhead compared to conventional FL. Numerical experiments show that the proposed framework exhibits higher spectral efficiency than the state-of-the-art SPIM-MIMO methods and mmWave-MIMO, which relies on the strongest propagation path.		Furthermore, the proposed FL approach provides at least $10$ times lower transmission overhead than the centralized learning techniques.
	\end{abstract}

	\section{Introduction}
	\label{sec:Introduciton}
	The millimeter wave multiple-input multiple-output (mmWave-MIMO) communications systems substantially improve the throughput in the fifth generation (5G) networks~\cite{mimoOverview,mimoScalingUp}.	As an emerging 5G technology, index modulation (IM) is attractive primarily because it offers both improved energy efficiency and spectral efficiency over conventional modulations. The IM encodes additional information in the indices of the transmission media such as subcarriers~\cite{im_eBasar_ComMag,hodge2020intelligent}, antennas~\cite{antenna_grouping_SM,hodge2019reconfigurable}, and spatial paths~\cite{spim_bounds_JSTSP,spim_BIM_TVT,spim_GBM}. In this paper, we focus on spatial modulation (SM) in the context of mmWave-MIMO systems~\cite{mishra2019toward}.
	
	In mmWave-MIMO, hybrid analog-digital beamformers are employed, where the number of radio-frequency (RF) chains is much smaller than the antennas. While this saves cost and power, its multiplexing gain is limited~\cite{mimoOverview}. The SM techniques have been shown to be helpful in addressing this problem~\cite{spim_bounds_JSTSP,spim_BIM_TVT,spim_GBM}. %\textcolor{red}{reference?}
	In~\cite{antenna_grouping_SM}, an antenna grouping (AG) approach is proposed for point-to-point communication, wherein some antenna elements are (de)activated to provide SM in terms of active/passive antenna indices. This approach suffers from reduced array gain because it always uses a subarray. For a single-user scenario, \cite{spim_BIM_TVT} 
	% and \cite{spim_GBM}
	proposed beamspace-based approaches for spatial path index modulation (SPIM), which modulates the indices of the spatial paths to create different \emph{spatial patterns}. The use of beamspace model is further exploited in~\cite{spim_GBM} by employing lens arrays at both transmitter and receiver to improve the bit-error-rate (BER). 
	
	Apart from BER, spectral efficiency is utilized as a performance metric in~\cite{spim_bounds_JSTSP} for SPIM-based transmitter design. Here, theoretical conditions for SPIM-MIMO to outperform mmWave-MIMO are introduced. The SPIM structure in~\cite{spim_bounds_JSTSP} considers only analog beamformer design, for which the same baseband beamformers are used even if the structure of the analog beamformer is changed due to the selection of different spatial patterns. Analog-only beamformer design is also considered in~\cite{spim_multiuser_TCOM} for uplink multi-user scenario with codebook design. A joint design for analog and baseband precoders for SPIM is performed in~\cite{spim_analog_digital} by implementing zero-forcing baseband precoding and selecting the steering vectors as analog beamformer candidates. Similar to~\cite{spim_analog_digital}, \cite{spim_bounds_JSTSP,spim_BIM_TVT} also design the analog precoders with a predefined codebook of steering vectors, which entails a beam training task prior to  the precoder design. Most of the aforementioned works investigate the single-user scenario. Their extension to the multi-user case remains a challenge. Although \cite{spim_multiuser_TCOM} considered the uplink multi-user SPIM architecture, it included the codebook of analog-only beamformers at the user end. 
	
	In this paper, we design both analog and digital beamformers for a downlink multi-user scenario using model-based and model-free techniques. We leverage the optimality of the manifold optimization (MO)~\cite{hybridBFAltMin,elbir2020withoutCSI} for the model-based approach. Then, taking advantage of the model-free structure of learning-based methods~\cite{elbir2020cognitive,elbir2020FL_HB,elbir2020_FL_CE} to improve robustness and computational efficiency, we train a global model through federated learning (FL). All users contribute to the learning process by computing the model updates with respect to their local datasets. The model updates are then collected at the base station (BS) for model aggregation and then sent back to the users for the next communication round and the global model is iteratively updated. Once trained, the  model parameters are shared with each user, which can estimate the beamformers by simply feeding the model with its downlink channel matrix. As a result, a non-linear data mapping is constructed between the channel data (input) and the beamformers (output), wherein a convolutional neural network (CNN) with dropout learning (DL) is designed~\cite{caldas2018expanding}. The DL allows randomly selecting a fraction (up to \texttildelow$50\%$) of the model parameters, thus further reducing the communication cost during FL-based training.

	Unlike the conventional centralized learning (CL) methods~\cite{elbirQuantized_TWC_2020,elbirHybrid_multiuser,mimoDeepPrecoderDesign}, where the BS collects all of the training datasets from users, our proposed FL-based approach is advantageous because of less transmission overhead; it is further reduced by employing DL to send approximately half of the model parameters to the users. We validate this through extensive numerical experiments and demonstrate that the proposed model-based and model-free approaches have superior spectral efficiency than the state-of-the-art model-based SPIM techniques~\cite{spim_bounds_JSTSP} as well as outperforming the conventional mmWave-MIMO~\cite{mimoHybridLeus3}. Apart from maintaining satisfactory prediction performance, the model-free FL offers a communication-efficient training, which requires approximately $10$ times lower communication exchange for model parameter transmission than the conventional CL-based techniques.

	\section{System Model}
\begin{figure}[t]
		\centering		{\includegraphics[draft=false,width=.8\columnwidth]{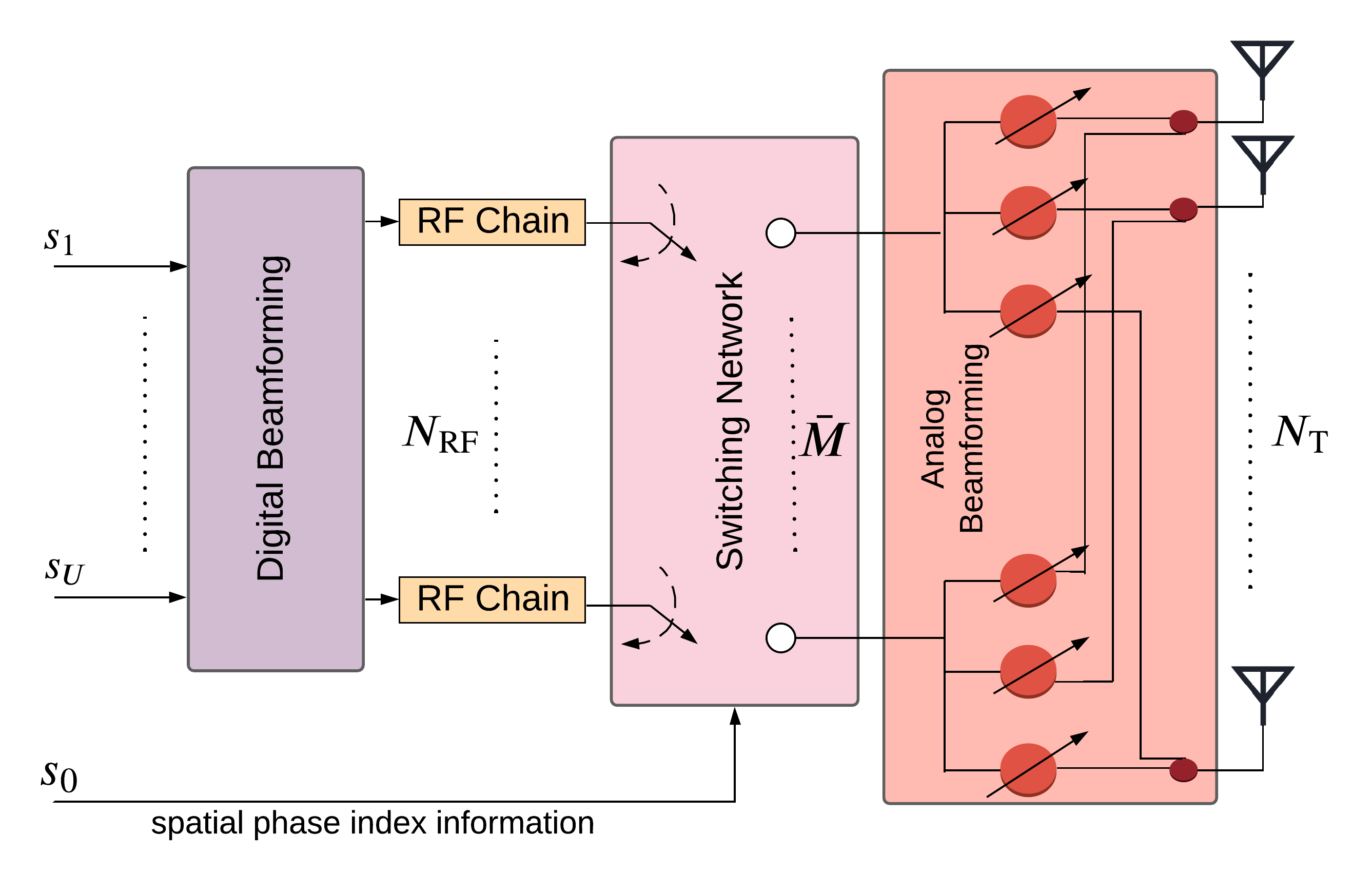} }
		\vspace*{-5mm} 
		\caption{The SPIM-MIMO architecture processes the incoming data streams $\{{s}_u\}_{u\in \mathcal{U}}$ and employs spatial path index information $s_0$ in a switching network, which connects $N_\mathrm{RF} = U$ RF chains to $\bar{M} = UM$ taps on the analog beamformers to exploit one of the $M$ spatial paths per user.
		}
		%			\vspace*{-5mm}
		\label{fig_BS}
	\end{figure}

	Consider a multi-user MIMO scenario with SPIM (SPIM-MIMO), where the BS has $N_\mathrm{T}$ antennas to communicate with $U$ users, each of which has $N_\mathrm{R}$ antennas, via a single data stream. Then, the vector of all data symbols are given by ${\mathbf{s}} = [{{s}}_1,\dots, {{s}}_U]^\textsf{T}\in \mathbb{C}^{U}$.  Additionally, the spatial path index information represented by ${s}_0$ is fed to the switching network (Fig.~\ref{fig_BS}) to randomly assign the outputs of $N_\mathrm{RF} = U\leq  \bar{M}$ RF chains to the $ \bar{M}$ taps of the analog beamformer. Thus, the BS can process at most $ \bar{M}$ spatial paths, for which $ \bar{M} = UM \leq N_\mathrm{T}$, where $M$ denotes the number of available spatial paths for each user. Compared to the conventional mmWave-MIMO, SPIM-MIMO has the advantage of transmitting additional data streams by exploiting the \emph{spatial pattern} of the mmWave channel with limited RF chains, i.e., $N_\mathrm{RF} \leq  \bar{M}$~\cite{spim_bounds_JSTSP}. If $ M=1$, i.e., $N_\mathrm{RF} = \bar{M}$, then SPIM-MIMO reduces to conventional mmWave-MIMO because there is only one choice of transmission~\cite{spim_GBM}.
	
	Assume $\mathbf{F}_\mathrm{RF}^{(i)}\in \mathbb{C}^{N_\mathrm{T}\times U}$ and $\mathbf{F}_\mathrm{BB}^{(i)}\in \mathbb{C}^{U\times U}$ be the analog and baseband beamformers corresponding to the $i$-th spatial pattern, respectively, for $i = 1,\dots, M^U$, i.e., selecting one of the $M$ paths for each user%\footnote{The assumption of selecting one out of $M$ paths can be generalized for SPIM with selecting more than one out of $M$~\cite{spim_bounds_JSTSP}, which we reserve to study for a future work.}
	. The signal vector transmitted by the $N_\mathrm{T}$ antennas is 
	\begin{align}
	\label{txSignal}
	\mathbf{x}^{(i)} = \mathbf{F}_\mathrm{RF}^{(i)} \mathbf{F}_\mathrm{BB}^{(i)}\mathbf{s}.
	\end{align}
	Note that (\ref{txSignal}) includes the design of both analog and baseband beamformers for each spatial pattern whereas the method in~\cite{spim_bounds_JSTSP} designs only analog beamformers and uses a fixed baseband beamformer.
	
	The RF precoders $\mathbf{F}_\mathrm{RF}^{(i)}$, which are constructed by phase shifters, have constant-modulus elements, i.e., $|[\mathbf{F}_\mathrm{RF}^{(i)}]_{m,n}| = \frac{1}{\sqrt{N_\mathrm{T}}} $. In addition,  we have the power constraint $\|\mathbf{F}_\mathrm{RF}^{(i)} \mathbf{F}_\mathrm{BB}^{(i)} \|_\mathcal{F}^2 $ $= N_\mathrm{RF}$ that is enforced by the normalization of $\mathbf{F}_\mathrm{BB}^{(i)}$. Finally, the $N_\mathrm{R}\times 1$ received signal by the $u$-th user becomes
	\begin{align}
	\label{receivedSignal1}
	\mathbf{y}_u^{(i)} = \mathbf{H}_u \mathbf{F}_\mathrm{RF}^{(i)} \mathbf{F}_\mathrm{BB}^{(i)}\mathbf{s} + \mathbf{n}_u,
	\end{align}
	where $\mathbf{H}_u\in \mathbb{C}^{ N_\mathrm{R}\times N_\mathrm{T}}$ represents the mmWave channel matrix between the BS and the $u$-th user and $\mathbf{n}_u \sim \mathcal{CN}(0,\sigma_n^2\mathbf{I}_{N_\mathrm{R}})$ is temporarily and spatially white zero-mean Gaussian noise with variance $\sigma_n^2$. The mmWave channel can be modeled as the contribution of $M$ clustered paths from each user~\cite{mimoRHeath,mimoHybridLeus3}. Thus, $\mathbf{H}_u$ can be given by
	\begin{align}
	\mathbf{H}_u = \mathbf{A}_\mathrm{R}^{(u)}\boldsymbol{\Sigma}_u \mathbf{A}_\mathrm{T}^{(u)^\textsf{H}},
	\end{align}
	where the steering matrices $\mathbf{A}_\mathrm{R}^{(u)} = [\mathbf{a}_\mathrm{R}^{(u)}(\phi_1),\dots,\mathbf{a}_\mathrm{R}^{(u)}(\phi_M)]$ $\in \mathbb{C}^{N_\mathrm{R}\times M}$
	and  $\mathbf{A}_\mathrm{T}^{(u)} = [\mathbf{a}_\mathrm{T}^{(u)}(\varphi_m),\dots,\mathbf{a}_\mathrm{T}^{(u)}(\varphi_M)]\in \mathbb{C}^{N_\mathrm{T}\times M}$ correspond to the 
	angle-of-arrival/angle-of-departure (AoA/AoD) angles $\phi_m$ and $\varphi_m$, for $m = 1,\dots, M$, respectively. For a uniform linear array (ULA), the $n$-th element of $\mathbf{a}_\mathrm{R}^{(u)}(\phi)$ and $\mathbf{a}_\mathrm{T}^{(u)}(\varphi)$ can be defined as $[\mathbf{a}_\mathrm{R}^{(u)}( \phi)]_n  = \frac{1}{\sqrt{N_\mathrm{R}}}\exp\{-\mathrm{j}\pi $ $ (n-1) \sin (\phi) \}$ and $[\mathbf{a}_\mathrm{T}^{(u)}( \varphi)]_n =\frac{1}{\sqrt{N_\mathrm{T}}}\exp\{-\mathrm{j}\pi $ $ (n-1) \sin (\varphi) \}$, respectively. $\boldsymbol{\Sigma}_u = \mathrm{diag}\{\sqrt{\gamma_{u,1}},\dots, \sqrt{\gamma_{u,M}}  \}$ is an $M\times M$ diagonal matrix including the scattering path gains $\gamma_{u,m}$~\cite{spim_bounds_JSTSP}.
	
	The received signal $\mathbf{y}_u^{(i)}$ is then processed by analog combiner  $\mathbf{w}_\mathrm{RF}^{(u,i)}\in \mathbb{C}^{N_\mathrm{R}}$
	%	 and the signal to be detected in the baseband can given by $\widetilde{{y}}_u^{(i)} =  \mathbf{w}_\mathrm{RF}^{(u,i)^\textsf{H}}\mathbf{y}_u^{(i)}$
	as
	\begin{align}
	\widetilde{{y}}_u^{(i)} =  \mathbf{w}_\mathrm{RF}^{(u,i)^\textsf{H}}\mathbf{H}_u \mathbf{F}_\mathrm{RF}^{(i)} \mathbf{F}_\mathrm{BB}^{(i)}\mathbf{s} + \widetilde{{n}}_u,
	\end{align}
	where $\widetilde{{n}}_u =  \mathbf{w}_\mathrm{RF}^{(u,i)^\textsf{H}}\mathbf{n}_u$. Similar to the analog precoders, the analog combiner $\mathbf{w}_\mathrm{RF}^{(u,i)}$ also has constant-modulus elements, i.e., $|[\mathbf{w}_\mathrm{RF}^{(u,i)}]_{n}| = {1}/{\sqrt{N_\mathrm{R}}} $, $n = 1,\dots, N_\mathrm{R}$.

	%%-----------------------------------------------------
	%%-----------------------------------------------------

	Our goal is to design the  beamformers, $\mathbf{F}_\mathrm{RF}^{(i)}$, $\mathbf{F}_\mathrm{BB}^{(i)}$ and $\mathbf{W}_\mathrm{RF}^{(i)} = [\mathbf{w}_\mathrm{RF}^{(1,i)},\dots,\mathbf{w}_\mathrm{RF}^{(U,i)}]$ by exploiting SPIM. The downlink channel $\mathbf{H}_u$ is available for $u\in \mathcal{U} = \{1,\dots,U\}$ and used to design the beamformers with FL-based training, in which a  learning model is trained to provide a  mapping from the channel matrix to the beamformers.  %In the following, we first present a model-based approach to construct the analog beamformers for all possible $M$ paths for each users.
	%	, i.e. $\overline{\mathbf{F}}_\mathrm{RF}^{(u)}\in  \mathbb{C}^{N_\mathrm{T}\times M}$. 
	%Next, the solution obtained from the model-based approach is used as a label in the training data to train a model via FL.

	\section{ Beamforming via Model-based Approach}
	We construct the analog beamformers via simultaneously incorporating all of the $M$ spatial paths, in which the analog beamformers corresponding to all spatial paths $\overline{\mathbf{F}}_\mathrm{RF}^{(u)}= [\overline{\mathbf{f}}_\mathrm{RF}^{(u,1)},\dots, \overline{\mathbf{f}}_\mathrm{RF}^{(u,M)} ] \in  \mathbb{C}^{N_\mathrm{T}\times M}$  per user and $\overline{\mathbf{W}}_\mathrm{RF}^{(u)}= [\overline{\mathbf{w}}_\mathrm{RF}^{(u,1)},\dots, \overline{\mathbf{w}}_\mathrm{RF}^{(u,M)} ]$ $ \in \mathbb{C}^{N_\mathrm{R}\times M}$ are designed for $u\in \mathcal{U}$. Then, we design the baseband precoders after taking into account the interference among the users.  Given $\mathbf{H}_u$\footnote{The estimate of $\mathbf{H}_u$ is obtained via both learning-~\cite{elbir2019online,elbir2020_FL_CE,elbir_LIS} and non-learning-based~\cite{mimoHybridLeus3,widebandChannelEst2} approaches. We assume  $\mathbf{H}_u$ is obtained prior to the beamformer design.}, the analog beamformer $\overline{\mathbf{F}}_\mathrm{RF}^{(u)}\in  \mathbb{C}^{N_\mathrm{T}\times M}$ is designed by minimizing the distance between the beamformer $\overline{\mathbf{F}}_\mathrm{RF}^{(u)}\overline{\mathbf{f}}_\mathrm{BB}^{(u)} \in  \mathbb{C}^{N_\mathrm{T}}$  and the optimal digital precoder $\mathbf{f}_u^\mathrm{opt}\in \mathbb{C}^{N_\mathrm{T}}$, available from singular value decomposition (SVD) of $\mathbf{H}_u$~\cite{mimoRHeath}. Thus, the maximizing the spectral efficiency~\cite{elbirHybrid_multiuser} is equivalent to solve 
	%, $\overline{\mathbf{F}}_\mathrm{BB}\in  \mathbb{C}^{M\times K}$ 
	\begin{align}
	\label{designFrf}
	%\{ 
	%	 \overline{\mathbf{F}}_\mathrm{RF}^{(u)}
	%, \overline{\mathbf{F}}_\mathrm{BB}^{(u)}\}
	%	 = &\arg
	\minimize_{\overline{\mathbf{F}}_\mathrm{RF}^{(u)}, \overline{\mathbf{f}}_\mathrm{BB}^{(u)}} &\;\; \|\mathbf{f}_u^\mathrm{opt} - \overline{\mathbf{F}}_\mathrm{RF}^{(u)}\overline{\mathbf{f}}_\mathrm{BB}^{(u)}   \|_\mathcal{F}^2 \nonumber   \\
	\subjectto & \;\;|[\overline{\mathbf{F}}_\mathrm{RF}^{(u)}]_{m,n}| = \frac{1}{\sqrt{N_\mathrm{T}}},
	\;\; \|\overline{\mathbf{F}}_\mathrm{RF}^{(u)} \overline{\mathbf{f}}_\mathrm{BB}^{(u)} \|_\mathcal{F}^2 = M,
	\end{align}
	which is solved for $u\in \mathcal{U}$ to obtain the analog precoders $\{\overline{\mathbf{F}}_\mathrm{RF}^{(u)}\}_{u\in \mathcal{U}}$. 
	
	Similarly, the following optimization yields analog combiners $\overline{\mathbf{W}}_\mathrm{RF}^{(u)}\in \mathbb{C}^{N_\mathrm{R}\times M} $ for all possible paths:
	\begin{align}
	\label{designWrf}
	&\underset{\overline{\mathbf{W}}_\mathrm{RF}^{(u)}, \overline{\mathbf{w}}_\mathrm{BB}^{(u)}}{\minimize}
	\;\; \| {\mathbf{w}}_\mathrm{MMSE}^{(u)}
	- \overline{\mathbf{W}}_\mathrm{RF}^{(u)} \overline{\mathbf{w}}_\mathrm{BB}^{(u)} \|_\mathcal{F}^2 \nonumber \\
	&\subjectto  \;\;|[\overline{\mathbf{W}}_\mathrm{RF}^{(u)}]_{n,m}| = \frac{1}{\sqrt{N_\mathrm{R}}},
	\end{align}
	where ${\mathbf{w}}_\mathrm{MMSE}^{(u)^\textsf{H}} =\big( \mathbf{f}_u^{\mathrm{opt}^\textsf{H}}\mathbf{H}_u^\textsf{H}\mathbf{H}_u \mathbf{f}_u^\mathrm{opt}	+ {\sigma_n^2} \big)^{-1}\mathbf{f}_u^{\mathrm{opt}^\textsf{H}}\mathbf{H}_u^\textsf{H}$ is the $N_\mathrm{R}\times 1$ optimum combiner using minimum-mean-squared-error (MMSE) estimation, which is used to obtain unconstrained combiner~\cite{mimoRHeath}. $\overline{\mathbf{w}}_\mathrm{BB}^{(u)}= (\overline{\mathbf{W}}_\mathrm{RF}^{(u)^\textsf{H}} \boldsymbol{\Lambda}_{\mathrm{y}_u} \overline{\mathbf{W}}_\mathrm{RF}^{(u)})^{-1} (\overline{\mathbf{W}}_\mathrm{RF}^{(u)^\textsf{H}}\boldsymbol{\Lambda}_{\mathrm{y}_u}{\mathbf{w}}_\mathrm{MMSE}^{(u)}) $ $\in \mathbb{C}^{M}$ is used to compute all analog combiners. Once $\overline{\mathbf{W}}_\mathrm{RF}^{(u)}$ is found, the receiver only uses a single column of $\overline{\mathbf{W}}_\mathrm{RF}^{(u)}$ as a combiner vector.  The covariance matrix of the received signal in (\ref{receivedSignal1}), for which the analog and baseband precoders are replaced with $\overline{\mathbf{F}}_\mathrm{RF}^{(u)}$ and $\overline{\mathbf{F}}_\mathrm{BB}^{(u)}$, is $\boldsymbol{\Lambda}_{\mathrm{y}_u} = \mathbf{H}_u\overline{\mathbf{F}}_\mathrm{RF}^{(u)}\overline{\mathbf{F}}_\mathrm{BB}^{(u)}  \overline{\mathbf{F}}_\mathrm{BB}^{(u)^\textsf{H}}\overline{\mathbf{F}}_\mathrm{RF}^{(u)^\textsf{H}}\mathbf{H}_u^\textsf{H} $ $+ \sigma_n^2\mathbf{I}_{N_\mathrm{R}}\in \mathbb{C}^{N_\mathrm{R}\times N_\mathrm{R}}$.
	
	The optimization problems in (\ref{designFrf}) and (\ref{designWrf}) are effectively solved via alternating minimization techniques, such as manifold optimization (MO) or ``Manopt'' algorithm~\cite{hybridBFAltMin,elbir2019online,elbir2020withoutCSI}. 
	%		Although the Manopt algorithm is computationally complex as compared to the conventional greedy search~\cite{mimoRHeath,alkhateeb2016frequencySelective} and phase extraction-based~\cite{sohrabiOFDM} approach, it yields optimal beamforming results, which are used as labels of the proposed learning model. 
	This is optimal in the sense that it achieves the minimum Euclidean distance between the unconstrained and hybrid beamformers.

	To exploit SPIM, only one column of $\overline{\mathbf{F}}_\mathrm{RF}^{(u)}$ and $\overline{\mathbf{W}}_\mathrm{RF}^{(u)}$ is selected for the $i$-th spatial pattern as ${\mathbf{f}}_\mathrm{RF}^{(u,i)} =\overline{\mathbf{F}}_\mathrm{RF}^{(u)}\mathbf{b}^{(u,i)} $ and ${\mathbf{w}}_\mathrm{RF}^{(u,i)} = \overline{\mathbf{W}}_\mathrm{RF}^{(u)}\mathbf{b}^{(u,i)}$.  Denote $\mathcal{B}^{(u)} =\{b^{(u,i_1)},\dots,b^{(u,i_M)} \} $ to be the set of selected paths for all possible path configurations of the $u$-th user for $i_m = 1,\dots, M$. Then, the entries of $\mathbf{b}^{(u,i)} \in \mathbb{R}^{M}$ are all zeros except the $i_m$-th element, which is unity and denotes selection of the $i_m$-th spatial path for the $u$-th user.

	In order to mitigate interference among the users, the baseband beamformer needs to be updated by computing the effective channel as $\footnotesize\mathbf{H}_\mathrm{eff}^{(i)} = \left(\begin{array}{c}	\mathbf{h}_\mathrm{eff}^{(1,i)}\\ \vdots \\ \mathbf{h}_\mathrm{eff}^{(U,i)}	\end{array}\right) \in \mathbb{C}^{U\times U}$, where $\mathbf{h}_\mathrm{eff}^{(u,i)} = {\mathbf{w}}_\mathrm{RF}^{(u,i)} \mathbf{H}_u {\mathbf{F}}_\mathrm{RF}^{(i)} \in \mathbb{C}^{1\times U}  $ where $ {\mathbf{F}}_\mathrm{RF}^{(i)} = [ {\mathbf{f}}_\mathrm{RF}^{(1,i)},\dots,{\mathbf{f}}_\mathrm{RF}^{(U,i)}]$.  Finally, the baseband precoder ${\mathbf{F}}_\mathrm{BB}^{(i)}$ is obtained as ${\mathbf{F}}_\mathrm{BB}^{(i)} = \mathbf{H}_\mathrm{eff}^{(i)^{-1}}$ and it is normalized as $[{\mathbf{F}}_\mathrm{BB}^{(i)}]_{u,:} = {[{\mathbf{F}}_\mathrm{BB}^{(i)}]_{u,:}}/{\|  [{\mathbf{F}}_\mathrm{BB}^{(i)}]_{u,:}  \|_2 }$.

	\section{ Beamforming  via Federated Learning}
	%This section presents our FL approach to design the beamformers. 
	The learning model accepts $\mathbf{H}_u$ as input and yields ${\mathbf{F}}^{(i)} = \mathbf{F}_\mathrm{RF}^{(i)}\mathbf{F}_\mathrm{BB}^{(i)}\in \mathbb{C}^{N_\mathrm{T}\times U}$ and ${\mathbf{w}}_\mathrm{RF}^{(u,i)}$ at the output.
	Define $\mathcal{D}_u$ be the local dataset of the $u$-th user, in which the $l$-th element is $\mathcal{D}_l = (\mathcal{X}_u^{(l)},\mathcal{Y}_u^{(l)})$, where  $\mathcal{X}_u^{(l)}$ and $\mathcal{Y}_u^{(l)}$ are the input and output  for $l =1,\dots,\textsf{D}_u$, and $\textsf{D}_u = |\mathcal{D}_u|$ is the size of the local dataset.	The input $\mathcal{X}_u\in \mathbb{R}^{N_\mathrm{R}\times N_\mathrm{T}\times 3}$ can be constructed by ``three-channel'' data, whose the first and second ``channel'' can be designed as the element-wise real and imaginary part of $\mathbf{H}_u$ as  $[\mathcal{X}_u]_1 = \operatorname{Re}\{\mathbf{H}_u \}$ and $[\mathcal{X}_u]_2 = \operatorname{Im}\{\mathbf{H}_u \}$, respectively. Also, the third channel can be constructed as $[\mathcal{X}_u]_3 = \angle \{\mathbf{H}_u \}$, which is demonstrated to improve the feature extraction performance~\cite{elbirDL_COMML,elbirQuantized_TWC_2020}. Then, the output $\mathcal{Y}_u\in \mathbb{R}^{(2N_\mathrm{T}U + N_\mathrm{R})\times 1}$ is constructed as $\mathcal{Y}_u = \mathrm{vec}\{\mathrm{vec}\{\operatorname{Re}\{ \mathbf{F}^{(i)} \}, \operatorname{Im}\{ \mathbf{F}^{(i)} \}    \}^\textsf{T},  \angle\mathbf{w}_\mathrm{RF}^{(u,i)^\textsf{T}} \}^\textsf{T}     $.
	
	In FL, the training dataset $\mathcal{D}$ is partitioned into small portions, i.e., $\mathcal{D}_u$, $u\in \mathcal{U}$, which are available at the users and not transmitted to the BS. Let $\boldsymbol{\theta}\in \mathbb{R}^{P}$  denote the learnable parameters of size $P$, then FL solves the following problem for the $t$-th communication round of the model training, i.e., $\minimize_{\boldsymbol{\theta}}    
	\frac{1}{\textsf{D}_u} \sum_{l = 1}^{\textsf{D}_u}\mathcal{L}(f( \mathcal{X}_u^{(l)}|\boldsymbol{\theta}_{t-1}),\mathcal{Y}_u^{(l)}  ),$
	with the use of the local gradient $\mathbf{g}_u(\boldsymbol{\theta}_t)$, where $\boldsymbol{\theta}_t$ denotes the model parameters at the $t$-th iteration and $\mathcal{L}(\cdot)$ is the loss function. Then, the $u$-th user transmits $\mathbf{g}_u(\boldsymbol{\theta}_t)$ to the BS. Once the gradient data from all users are collected, the BS finally incorporates ${\mathbf{g}}_u(\boldsymbol{\theta}_t)$ for $u\in \mathcal{U}$ to update $\boldsymbol{\theta}_t$ as $\boldsymbol{\theta}_{t+1} = \boldsymbol{\theta}_t - \eta_t  \frac{1}{U} \sum_{u=1}^{U} {\mathbf{g}}_u(\boldsymbol{\theta}_t),$
	where ${\mathbf{g}}_u(\boldsymbol{\theta}_t) = \frac{1}{\textsf{D}_u} \sum_{l=1}^{\textsf{D}_u}\nabla_{\boldsymbol{\theta}} \mathcal{L} (f ( \mathcal{X}_u^{(l)}|\boldsymbol{\theta}_t), \mathcal{Y}_u^{(l)})  )$ for learning rate $\eta_t$.
	After model aggregation, the BS returns the updated model parameters $\boldsymbol{\theta}_{t+1}$ to the users,	which will be used for the computation of the gradients in the next iteration. 
	
	% \subsection{Network Architecture}
	The proposed network architecture is a CNN comprised of $10$ layers. The first layer is the input layer, which accepts the input data of size $N_\mathrm{R}\times N_\mathrm{T}\times 3$.  The $\{2,4,6\}$-th layers are the convolutional layers with $N_\mathrm{SF} = 128$ filters, each of which employs a $3\times 3$ kernel for 2-D spatial feature extraction. The $\{3,5,7\}$-th layers are the normalization layers. The eighth layer is a fully connected layer with $N_\mathrm{FCL}=1024$ units, whose main purpose is to provide feature mapping. The ninth layer is a dropout layer with $\kappa=1/2$ probability. The dropout layer applies an $N_\mathrm{FCL}\times 1$ mask on the weights of the fully connected layer, whose elements are uniform randomly selected from $\{0,1\}$. As a result, at each iteration, DL randomly selects different set of weights in the fully connected layer, thereby reducing the size of $\boldsymbol{\theta}_t$ and $\mathbf{g}_u(\boldsymbol{\theta}_t)$, thereby, reducing model transmission overhead. Finally, the last layer is output regression layer, yielding the output channel estimate of size  $(2N_\mathrm{T}U+ N_\mathrm{R})\times 1$. Once the training is completed, each user feeds the model with $\mathbf{H}_u$ and obtains its beamformer $\mathbf{w}_\mathrm{RF}^{(u,i)}$ and $\mathbf{F}^{(i)}$, which is fed back to the BS.

	%\section{Transmission Overhead and Complexity Analysis}
	%\label{sec:Complexity}
	%\subsection{Transmission Overhead}
	\label{sec:DataComp}
	We further examine the transmission overhead which can be defined as the size of the transmitted data during model training. Let $\mathcal{T}_\mathrm{FL}$ and $\mathcal{T}_\mathrm{CL}$ denote the transmission overhead of FL and CL, respectively. Define $\textsf{D} = \sum_{u\in \mathcal{U}} \textsf{D}_u $ so that $\mathcal{T}_\mathrm{CL} = 
	\big(3N_\mathrm{T}N_\mathrm{R} + 2N_\mathrm{T}U+N_\mathrm{R}    \big) \textsf{D}, $ which includes the number of symbols in the uplink transmission of the training dataset $\mathcal{D}$ from the users to the BS. In contrast, the transmission overhead of FL includes the transmission of $\mathbf{g}_u(\boldsymbol{\theta}_t)$ and $\boldsymbol{\theta}_t$ in uplink and downlink communication for $t = 1,\dots,T$, respectively. Finally, $\mathcal{T}_\mathrm{FL}$ is given by $\mathcal{T}_\mathrm{FL} = 2PTU.$ We can see that the dominant terms are $\textsf{D}$ and $P$, which are the number of training data pairs and the number of CNN parameters, respectively. While $\textsf{D}$ can be adjusted according to the amount of available data at the users, $P$ is usually unchanged during model training.  Here, $P = {N_\mathrm{CL}(CN_\mathrm{SF} W_x W_y)} +   {\kappa N_\mathrm{SF}  W_x W_yN_\mathrm{FCL}  },$ where  $N_\mathrm{CL}=3$ is the number of convolutional layers and $C=3$ is the number of spatial ``channels''.  $W_x=W_y=3$ are the 2-D kernel sizes. As a result, we have $P=600,192$ whereas $P = 1,190,016$ if dropout layer is removed.

	%%-----------------------------------------------------
	\begin{figure}[t]
		\centering
		{\includegraphics[draft=false,width=\columnwidth]{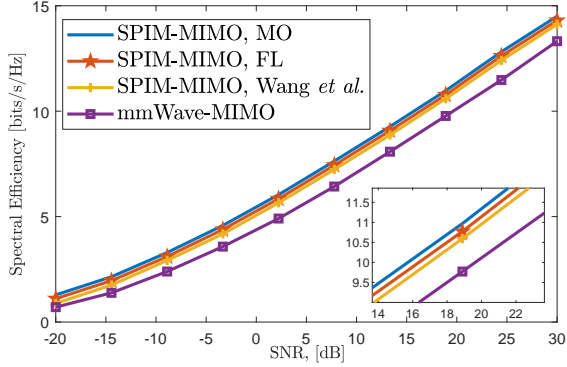} } 
		\vspace*{-6mm}
		\caption{Spectral efficiency for mmWave-MIMO and SPIM-MIMO.
		}
		
		\label{fig_SE1}
	\end{figure}
	%%-----------------------------------------------------

	\section{Numerical Simulations}
	\label{sec:Sim}
	We compared the performance of FL-based SPIM-MIMO with mmWave-MIMO and the state-of-the-art model-based SPIM-MIMO \emph{Wang et al.}~\cite{spim_bounds_JSTSP} in terms of spectral efficiency averaged over $1000$ Monte Carlo trials. 
	%	 During model training, the validation RMSE is defined by $	 \mathrm{RMSE} =\left(\frac{1}{|\mathcal{D}_\mathrm{val}|}  \sum_{l = 1}^{|\mathcal{D}_\mathrm{val}|}\| f( \widetilde{\mathcal{X}}^{(l)}|\boldsymbol{\theta}) - \widetilde{\mathcal{Y}}^{(l)}  \|_\mathcal{F}^2\right)^{1/2},$	where $\widetilde{\mathcal{X}}^{(l)}$ and $\widetilde{\mathcal{Y}}^{(l)}$ respectively denote the input-output pairs in the validation dataset $\mathcal{D}_\mathrm{val}$, which includes $20\%$ of the whole dataset $\mathcal{D}$, hence, we have $|\mathcal{D}_\mathrm{val}| = 0.2 |\mathcal{D}|$. 
	The local dataset of each user includes $N=200$ different channel realizations for $U=8$ users. The number of antennas at the BS and the users are $N_\mathrm{T}=128$ and $N_\mathrm{R}=9$, respectively. We select the number of available spatial paths for each user as $M=2$. The location of each user is selected as $\phi_{u,m}\in \Phi_u$ and  $ {\varphi}_{u,m} \in \bar{\Psi}_u$, for $m = 1,\dots, M$, where $\Phi_u$ and $\bar{\Psi}_u$ are the equally-divided subregions of the angular domain $\Theta =\bigcup_{u\in \mathcal{U}} \Phi_u = \bigcup_{u}\bar{\Psi}_u$, $\Theta\in [30^{\circ}, 150^{\circ}]$ as in~\cite{elbir2020FL_HB}. During training, each channel realization is corrupted by synthetic noise on the input data for three $\mathrm{SNR}_\mathrm{TRAIN}$ levels, i.e., $\mathrm{SNR}_\mathrm{TRAIN}=\{20, 25, 30\}$ dB,  for $G=200$ realizations in order to provide robust performance against noisy input~\cite{elbirDL_COMML,elbir_LIS}. As a result, the number of input-output pairs in the whole  training dataset is $\textsf{D}= 3UNG = 3\times8\times200\times200 = 960,000$. 
	
	The proposed CNN model is realized and trained in MATLAB on a PC with a $2304$-core GPU. For CL, we use the stochastic gradient descent (SGD) algorithm with momentum of $0.9$ and the mini-batch size $M_B = 128$,  and  update the network parameters with learning rate $0.001$. For FL, we train the CNN for $T=50$ iterations/rounds. Once the training is completed, the labels of the validation data (i.e., $20\%$ of the whole dataset) are used in prediction stage. 
	
	It was shown in~\cite{spim_bounds_JSTSP} that SPIM-MIMO outperforms mmWave-MIMO for $M=2$ with $\gamma_{1} \leq 4\gamma_{2}$, where $\gamma_m = \gamma_{u,m}$ for $u\in \mathcal{U}$ and $m =1,2$. Figure~\ref{fig_SE1} shows the spectral efficiency with respect to $\mathrm{SNR}$ when the spatial path gains for all users are selected as $\gamma_1 =\gamma_{2} = 0.5$. Note that both SPIM-MIMO and mmWave-MIMO use the same number of RF chains while SPIM-MIMO exploits the spatial distribution of the paths. In contrast, mmWave-MIMO designs the precoders in accordance to the largest path gains, i.e., $\gamma_{1}$, in our case. We observe that \emph{Wang et al.} provides less spectral efficiency than the proposed model-based approach because it employs a single baseband beamformer for all spatial patterns whereas the proposed model-based approach updates the baseband beamformer $\mathbf{F}_\mathrm{BB}^{(i)}$ in accordance to the different spatial patterns as well as suppressing the interference among the users. The proposed FL approach has slight performance loss than the model-based method due to the loss during model training. It is worth noting that the performance of FL is upper bounded
	by the model-based technique since FL cannot perform better than its labels.
	
\begin{figure}[t]
		\centering
		{\includegraphics[draft=false,width=\columnwidth]{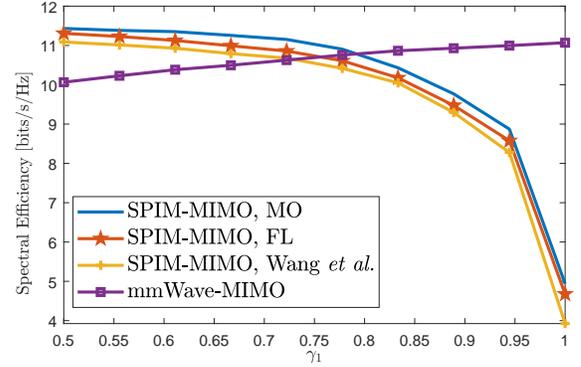} } 
		\vspace*{-6mm}
		\caption{Spectral efficiency versus $\gamma_{1}$ when $\mathrm{SNR} = 20$ dB.
		}
		
		\label{fig_SE_w1}
	\end{figure}

	In Fig.~\ref{fig_SE_w1}, we compare SPIM-MIMO and mmWave-MIMO with respect to $\gamma_{1}$ when $\gamma_{2} = 1-\gamma_{1}$. We observe that both techniques meet when $\gamma_{1} = 4\gamma_{2}$ for $\gamma_{1} = 0.8$. This clearly shows that the usage of SPIM is appropriate if the path gain are close. The SPIM-MIMO performance degrades as long as the difference between the path gains are large. As a result, mmWave-MIMO becomes favorable. We note from both Fig.~\ref{fig_SE1} and Fig.~\ref{fig_SE_w1} that our proposed FL approach closely follows the model-based technique.

	%%-----------------------------------------------------
	%%-----------------------------------------------------

	%%-----------------------------------------------------
	\begin{figure}[t]
		\centering
		{\includegraphics[draft=false,width=\columnwidth]{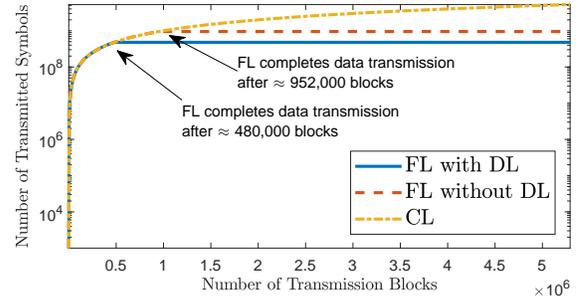} } 
		\vspace*{-4mm}
		\caption{Transmission overhead comparison of CL with FL, including and excluding DL.
		}
		
		\label{fig_TO}
	\end{figure}
	%%-----------------------------------------------------

	Next, we present the effectiveness of FL-based model training by comparison to the CL-based training.  According to the analysis in Sec.~\ref{sec:DataComp}, the transmission overhead of FL and CL are $2PTU = 2\cdot 600,192 \cdot 50 \cdot 8 \approx 480\times 10^6$ and $ (3N_\mathrm{T} N_\mathrm{R} + 2N_\mathrm{T}U + N_\mathrm{R}) \textsf{D} = (3\cdot 128\cdot 9 + 2\cdot128\cdot  8 + 9)\cdot 960,000 \approx 5.3\times 10^{9}$, respectively. This clearly shows the effectiveness of FL over CL, i.e., approximately $10$ times lower transmission overhead. In Fig.~\ref{fig_TO}, we visualize the number of transmitted symbols with respect to transmission blocks, each of which carries $1000$ symbols. We see that FL completes model training quicker than CL after approximately $480,000$ and $952,000$ transmission blocks with and without DL, respectively.

	\section{Summary}
	\label{sec:Conc}
	We presented both model-based and model-free frameworks for beamformer design in multi-user SPIM-MIMO systems. Whereas the former leverages MO for beamformer design, the latter employs FL to efficiently train the learning model. Our experiments showed that the proposed approach has superior performance than the state-of-the-art SPIM techniques as well as outperforming the conventional mmWave-MIMO systems in terms of spectral efficiency. Furthermore, the proposed FL approach exhibits a more communication-efficient learning method than conventional CL for model training. We demonstrated that FL with (without) DL enjoys approximately $10$ ($5$) times lower transmission overhead during model training lower transmission overhead than CL.

	%	 \textcolor{red}{Some more specific points needed in the summary. The last two sentences are same as abstract and Intro. Please change or quote some other result.} %As future work, we reserve to study for larger combination of spatial paths per user under the effect of model quantization and sparsification.

	%	\clearpage
	\newpage
	
	%	\balance
	\bibliographystyle{IEEEtran}
	\bibliography{IEEEabrv,references_070_journal,references_056_journal}

\end{document}